\documentclass[letterpaper]{article}
\usepackage{aaai}
\usepackage{times}
\usepackage{helvet}
\usepackage{courier}
\usepackage{multirow}
\usepackage{graphicx}
\usepackage[table,xcdraw]{xcolor}

\title{Visualising Generative Spaces Using Convolutional Neural Network Embeddings}

\author{Oliver Withington\\
Queen Mary University of London\\
London, UK\\
o.withington@qmul.ac.uk\\
\And
Laurissa Tokarchuk\\
Queen Mary University of London\\
London, UK\\
laurissa.tokarchuk@qmul.ac.uk\\}

\begin{document}
%

\maketitle
\begin{abstract}
\begin{quote}
As academic interest in procedural content generation (PCG) for games has increased, so has the need for methodologies for comparing and contrasting the output spaces of alternative PCG systems. In this paper we introduce and evaluate a novel approach for visualising the generative spaces of level generation systems, using embeddings extracted from a trained convolutional neural network. We evaluate the approach in terms of its ability to produce 2D visualisations of encoded game levels that correlate with their behavioural characteristics. The results across two alternative game domains, Super Mario and Boxoban, indicate that this approach is powerful in certain settings and that it has the potential to supersede alternative methods for visually comparing generative spaces. However its performance was also inconsistent across the domains investigated in this work, as well as it being susceptible to intermittent failure. We conclude that this method is worthy of further evaluation, but that future implementations of it would benefit from significant refinement.
\end{quote}
\end{abstract}

\section{Introduction}

Procedural Content Generation (PCG) for games, despite being a relatively new research field, has become a very active and diverse one, with an increasing volume of novel works being produced across numerous subdomains of PCG research. Whenever a PCG system involves stochastic elements in its generative process to produce diverse output, it is important to understand what outputs are possible from a given system, a concept often referred to as a PCG system’s ‘generative space’. This is important for both PCG researchers, as well as game designers. In commercial settings, designers ideally want to know that all possible artefacts that could be generated from a given PCG system are desirable for their purposes, and that an alternative system or configuration would not produce better artefacts. Similarly, PCG researchers often want to be able to credibly claim that a novel system or approach is a meaningful improvement over what was previously possible, and the qualities of the output are a valuable component of any comparison.

In this work’s domain of PCG systems focused on the generation of game levels, a common approach for understanding generative spaces is to produce simplified visualisations of system output, most commonly using Expressive Range Analysis (ERA) \cite{smith2010}. ERA is an approach for understanding generative spaces by visualising a sample of levels in terms of two of their behavioural characteristics (BCs), which are most commonly heuristics for aesthetic or gameplay related qualities. ERA is commonly used for both the qualitative understanding of PCG system output, as well as quantitatively to compare aspects of alternative generators such as their relative diversities of output. However, ERA has several weaknesses as a visualisation approach, some of which we argue can be addressed with the alternative method presented in this paper. Most relevantly, ERA only allows for visualisions of generative spaces in terms of two BCs while maintaining the readability of a 2D graph, and also requires that BCs are calculated for every new set of levels to be visualised.

In this paper we present a novel alternative method for producing two dimensional visualisations of generative systems using Convolutional Neural Networks (CNNs), a subtype of deep learning system which are widely used for image recognition tasks \cite{russakovsky2015a}. The basic operation of this approach is as follows. First, we train a CNN to predict the BCs of levels based on their structure. This CNN is then used to extract embeddings for each level we wish to visualise from the penultimate layer of the network. These embeddings are assembled and then compressed using principal component analysis (PCA), a dimensionality reduction algorithm, to represent each level using only two dimensions. These compressions can then be visualised on a 2D scatter-plot in which each point represents a level. The goal is that these 2D visualisations are similar to conventional ERA, except that distance between levels is closely correlated with their values for multiple BC values rather than just two. The extent to which this is the case is the focus of this paper’s experiments. The stronger the correlation the more credibly we can claim that we are able to realise the benefits of ERA without some of its weaknesses.

\section{Background and Related Work}\label{background}
\subsection{Expressive Range Analysis}\label{era}

The most widely used and influential approach for visualising the generative spaces of PCG systems is Expressive Range Analysis (ERA). It was introduced by Gillian Smith and Jim Whitehead in 2010 \cite{smith2010} as a way for the designers of procedural level generators to understand their generative spaces. 

Its basic mode of operation is appealingly simple. First a set of levels is generated to represent a generative space, and then annotated with BCs of interest (also often referred to as simply ‘metrics’  \cite{herve2021}). These BCs can be quantitative heuristics for aspects of the experience of playing a level, such as heuristics for difficulty \cite{smith2010,horn2014,jadhav2021a}, or they can be more abstract features such as how linear a level is \cite{smith2010}. The set of levels can then be visualised as a 2D scatter plot or heatmap, in which the levels location is specified by two selected BCs. 

There have been some evolutions and innovations in how ERA can be applied, such as Summerville’s work on visualising multiple BCs simultaneously using corner plots \cite{summerville2018} and Cook et als work on designing interactive  tools for conducting  ERA on a PCG system with tunable parameters                       \cite{cook2021}, but most commonly ERA is used in its original form and it is is still used in contemporary state-of-the-art PCG research  \cite{smith2018,jadhav2021a,alvarez2022}.

\subsection{Machine Learning and PCG}

This work is also inspired by the domain of Machine Learning-based PCG approaches, commonly collectively referred to as PCGML. Over the past decade there has been a proliferation of different approaches using ML to tackle different challenges within PCG. They have been used to achieve diverse goals such as: generating new levels from single training examples \cite{awiszus2020}; blending training from alternative games to generate content for unseen games \cite{jadhav2021a}; and powering AI level design partners \cite{guzdial2019}. A belief shared and reinforced by these works as well as this one, is that useful knowledge about game levels and their utility can be predicted from their representations using ML. 

The approach in this paper uses a specific type of Neural Network called a Convolutional Neural Network (CNN). They are most commonly used for image recognition and analysis, but have also seen success in PCG research focused on level generation \cite{wulff-jensen2018,volz2018a,irfan2019}. However the primary inspiration for this approach did not come from the domain of PCG research but instead from that of art analysis, specifically the work of \cite{gardini2021}. They used a CNN-based approach to produce 2D visualisations of sets of artistic works which they then aimed to order based on age, without the system having prior knowledge of the creation date. Their use of CNNs to create information rich 2D representations of sets of artistic content appeared directly relevant to the challenge of visualising PCG generative spaces, and led to the development of this project.

In this work we make use of a popular CNN architecture called VGG-16 \cite{simonyan2015a}, which was developed by Simonyan and Zisserman and was a winning entry in the ImageNet Large Scale Visual Recognition Challenge \cite{russakovsky2015a}.

\subsection{Level Generators and Benchmarks}\label{gensandbms}

In this work we make use of two open source level corpuses to experimentally assess our visualisation approach.

The Mario AI Benchmark is a widely used experimental platform which has helped to make Super Mario one of the most popular domains for both PCG research and game AI research more broadly. It was developed to support the 2009 Mario AI competition \cite{togelius2009MarioAI2010} but it has gone on to be used in numerous research projects, including contemporary research on novel PCG approaches \cite{awiszus2020,fontaine2021,sarkar2020,cernygreen2020}. The modern version of the benchmark, supported by Ahmed Khalifa et al \cite{khalifa2009}, contains nine sets of 1000 Mario levels produced by different level generators as part of Horn et als work to compare alternative PCG systems \cite{horn2014}.

The other source of visualisable level sets that we make use of is Boxoban, an open source version of Sokoban developed by \cite{boxobanlevels} to support their research into model-free planning \cite{guez2019}. As part of their work they released a large set of levels which were procedurally generated with the goal of having varying levels of challenge for reinforcement learning agents. It consists of over 1.5 million levels, each composed of a 10 by 10 grid with four goals. The levels are split into three sets: Medium, Hard and Unfiltered. The Medium and Hard sets were generated using the approach explained in \cite{guez2019}, and were sorted based on the ability of a trained agent to solve them. The Unfiltered set were generated using the approach of \cite{racaniere2017}, implemented by Guez et al, and were not separated by difficulty. 

\section{Approach}\label{approach}

In this section we present the main stages and steps involved in this approach for producing visualisations of sets of 2D game levels, as well as the process we used to investigate whether or not the visualisations contain useful information. The code used to create and validate the visualisations is available at github.com/KrellFace/Generative-Space-Compression.

\subsection{Stage 1: CNN Architecture and Training}

The first stage of the approach involves training a CNN to be able to predict level BCs based on the levels encoded structure. In this work we implement two alternative CNN architectures: VGG-16 and a simple 5 layer ‘Basic’ CNN to use as a comparison point for the larger and more complex VGG-16 implementation. In VGG-16’s original configuration, it used a softmax layer for its final layer to make class predictions for input images. However, as we are predicting BCs which are continuous numeric values we replace the softmax layer with a final dense layer. Apart from this change we use VGG-16 in its original configuration D form as described in \cite{simonyan2015}. The simplified ‘Basic’ CNN we implement only uses three convolutional layers interlaced with two max pooling layers. See Table \ref{table_CNNArch} for a more detailed description of the two architectures. 

\begin{table}[h]
\centering
\caption{CNN Architectures}
\label{table_CNNArch}
\begin{tabular}{|l|l|}
\hline
\textbf{Basic}            & \textbf{VGG-16}           \\ \hline
Conv-32                   & Conv-64 x 2               \\ \hline
MaxPool                   & MaxPool                   \\ \hline
Conv-64                   & Conv-128 x 2              \\ \hline
MaxPool                   & MaxPool                   \\ \hline
Conv-64                   & Cov-256 x 3               \\ \hline
FC-64         & MaxPool                   \\ \hline
FC-(BC Count) & Conv-512 x 3              \\ \hline
                          & MaxPool                   \\ \hline
                          & Conv-513 x 3              \\ \hline
                          & MaxPool                   \\ \hline
                          & FC-4096 x 2   \\ \hline
                          & FC-1000       \\ \hline
                          & FC-(BC Count) \\ \hline
\end{tabular}
\end{table}

\begin{figure}[ht]
\centering
\includegraphics[width=3.3in]{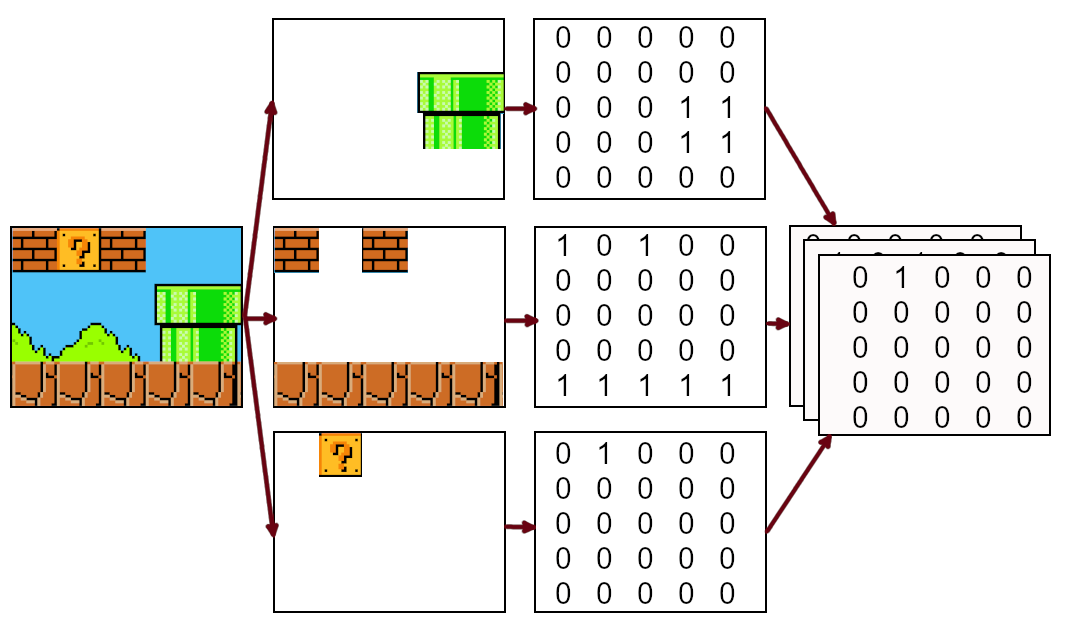}
\caption{Diagram showing how we generate a 3D one-hot matrix representation of game levels (Note: Only a subset of block types shown for brevity. Empty Tiles are included as a block type)}
\label{fig_onehot}
\end{figure}

To allow the CNN to process the levels, we first convert their character based representations into one-hot encoded representations. In the Boxoban and Mario AI Benchmark level corpuses each level is stored as a two dimensional matrix of characters, in which each character represents a specific tile that appears at that position. We take these 2D representations and convert them into a three dimensional one-hot matrix, with the size of the 3rd dimension equal to the number of possible tile types. The 3D matrix is effectively an assembled set of 2D matrices for each tile type, in which each value is either 0, or 1 in the case that a tile of that matrix’s type appears at that location (See Figure \ref{fig_onehot} for a visual explanation). This conversion to one-hot matrices is often used in PCGML, and was directly inspired by the work of Volz et al on their work using GANs to generate Mario levels \cite{volz2018a}.

For each training level, we first calculate its BCs and then process it into a one-hot 3D encoding. The CNN can then be trained to predict the BCs of unseen levels based on their one-hot encoded structures using this training data. Once training is completed the model is saved, ready for reuse in Stage 2.

\subsection{Stage 2: Embedding Extraction and Compression}

The next stage of the approach centres on using the trained neural network to produce level set visualisations in the form of 2D scatterplots, in which we intend that levels with similar BC values appear closer together in the visualisation than those with dissimilar characteristics. 

To accomplish this we take every level to be visualised, which in this paper’s experiments are levels not used in the original training, and process them with an instance of the previously trained CNN. However, rather than using the predicted BC values, we instead extract the output features from the penultimate fully connected dense layer from the CNN instance. In the case of VGG-16 this means extracting a 1D feature map of size 1000, and for the ‘Basic’ CNN of size 64. 

As we want to visualise these levels in two dimensional space, we then need to reduce the dimensionality of the data for each level. To achieve this we assemble the one dimensional feature maps for each level into a combined matrix, in which each row represents the penultimate layer outputs for a given level in the dataset to visualise. We then apply principal component analysis (PCA), a widely used dimensionality reduction algorithm to reduce the dimensionality of the dataset. PCA operates by constructing new variables out of linear combinations of the original variables in such a way that the top principal components account for the maximal possible variance in the data (See \cite{ayesha2020} for a more detailed explanation of PCA). We can then select the top two principal components and use them to represent the sets of levels using two dimensions, while still accounting for a substantial amount of the variance found in the original N dimensions extracted from the CNN. These levels can then be visualised on a standard 2D scatter plot, with their positions dictated by the principal components. 

\subsection{Stage 3: Validating the Visualisations}

The final stage of our approach is to validate whether the generated visualisations contain information that a human designer might find useful. In this work we use the same approach as  \cite{withington2022} and look for correlation between euclidean distances between levels in the visualisations, and the difference between their BC values. If strong correlation is found between the euclidean distances and the BC differences, then that means that levels with similar values for the BCs are found close together, and those with dissimilar BC values are further apart. 

To calculate the level of correlation we first calculate the euclidean distance between every level pair in the visualisation, as well as the absolute difference between their BC values. We then calculate Spearman’s rank correlation coefficient, also known as Spearman’s $\rho$. Spearman’s $\rho$ is commonly used to investigate the linear correlation between variables when the distribution is not expected to be normal. It uses the relative rankings of the data points rather than their values, to give a coefficient score from 0 to 1 on how strong the correlation is, along with a P value indicating the likelihood that the correlation found is actually present. These correlation coefficients are then used as the heuristic for the performance of that visualisation. The expectation is that the stronger the correlation the more useful the visualisations would be to a designer.

\begin{figure}[ht]
\centering
\includegraphics[width=3.3in]{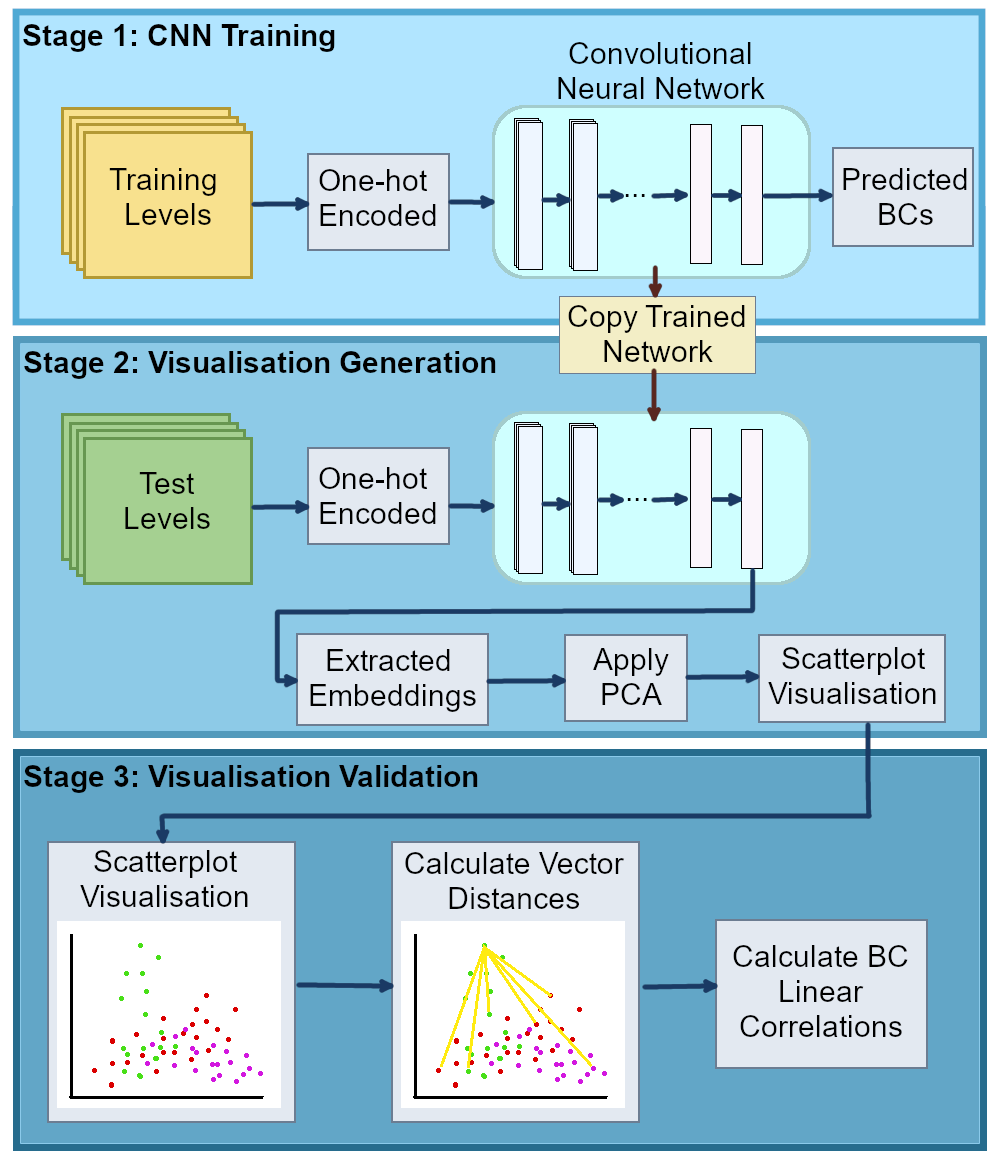}
\caption{Diagram showing the high level flow of all three stages of this generative space visualisation approach, as well as its validation in terms of BC correlation}
\label{fig_sysdiag}
\end{figure}
\section{Experiment Design}\label{design}

In this section we explain the design of the experiments that we ran to conduct an initial investigation of this approach. 

\subsection{Level Sets and Encodings}

For these experiments we use two pre-generated level corpuses: the level sets provided as part of the Mario AI Benchmark and the training levels provided as part of Guez et als research using Boxoban. Both corpuses meet several requirements for a straightforward application of the approach we present in this work. Both are composed of 2D tile-based levels in which every tile type is represented by one of a discrete set of values. Each level also has the same fixed size, 10 x 10 in the case of Boxoban and 16 x 100 in the case of the Mario AI Benchmark, which avoids the need  to regularise them before training the CNN. 

The two sets are also interestingly distinct. The Mario level sets come from a set of nine distinct generators with substantial variety in their structural and gameplay features (See both \cite{horn2014} and \cite{withington2022} for evidence of this). In contrast the Boxoban set all come from similar generative approaches implemented by the same team and are differentiated instead by difficulty, a trait that is less obvious than structural features. This combined with the smaller level representations, and the commonalities between all levels (all containing four boxes and four goals) mean it is significantly harder to differentiate between different levels. It is hoped that the difference of these two sets will help to highlight different strengths and weaknesses of the tested visualisation approaches.

In terms of the encoding used, in the case of Boxoban no preprocessing was required. Only five tile types are used in the level representation (empty, solid, box, goal, player) and we use all five in the one-hot encodings, giving a one-hot matrix with dimensions (10, 10, 5). In the case of the Mario level sets we conduct a preprocessing step of condensing the original tile types into a smaller grouped set. The original representations are composed of 28 total tile types, each of which we map to one of five values. All traversable tiles (including tiles that indicate the spawn point and level end point) are mapped to empty space and all enemies are mapped to a unifying enemy type. The same is done with solid blocks, pipe blocks and blocks that contain a reward. While this reduces the accuracy of the representations, it does increase the performance by reducing the complexity. The one hot representations are therefore of size (10, 16, 5).

\subsection{Behavioural Characteristics Used}\label{bcsused}

For both Mario and Boxoban domains we calculate four BCs for each level. Each BC is either one directly used in or inspired by prior work using ERA or Quality-Diversity (QD)-based PCG. QD-based PCG requires a designer to determine at least two behavioural dimensions to define the diversity component of the search, making it a valuable source of inspiration for BCs. Each BC was selected for both being fast to calculate from the level representation, as well as appearing in prior PCG work.

The four BCs calculated for the Mario domain were as follows: Empty Space (ES), a simple count of empty tiles, which was inspired by ERA implementations such as \cite{jemmali2020} which use block count-based BCs as quick to calculate heuristics for diversity. ES was calculated for both the Mario and Boxoban domains. Enemy Count (EC), Linearity (Lin) and Density (Dens) are all BCs which have been widely used in ERA applied to procedurally generated platformer levels \cite{smith2010,horn2014,jadhav2021a}, and make up the remaining three BCs calculated for the Mario levels. 

ERA and QD-search has been less commonly applied to top-down games like Boxoban than to platformer games like Super Mario. However, there are still sources to draw on. We calculate the Contiguity Score (CS), a BC which has been used in prior QD-search based PCG \cite{withington2020}, which measures how clustered together the solid tiles in a level are by incrementing the score by 1 for every solid block that is adjacent to another. We also calculate the Adjusted Contiguity Score (ACS), which divides the CS by the number of solid blocks present to give a heuristic for tile clustering independent of their number. Finally we calculate a Corridor Count (CC), which counts every location in the level in which the player can only move north or south but not east or west, or vice versa. This is intended to be a similar BC to Density except applicable to a top down game, and has appeared in work such as \cite{alvarez2022}.

\subsection{Compression Approaches Used}

As discussed in more detail in the Approach section, we implement two alternative CNN architectures to produce alternative visualisations, referred to in results as ‘VGG-16’ and ‘Basic’.

We also implement the approach of \cite{withington2022}, which works similarly to this paper’s CNN based approach except without the CNN processing layer. Instead of using PCA to compress the CNN embeddings, PCA is instead applied directly to the one-hot encoded levels. This approach makes for an interesting benchmark, as it both aims to generate similar 2D visualisations of generative spaces, while also requiring less setup and configuration in the form of calculating BCs and model training. We refer to this as ‘Vanilla DR’ in these experiments.

In this work we do not report on the explained variance of the top two principal components derived from applying PCA within any of the approaches as it is not relevant to the quality final output. However, prior reviews of this work have highlighted that explained variance could be a valuable metric to extract during CNN training for evaluating how effectively the model is learning to indirectly produce the desired visualisation, something we aim to investigate in future work. 

\subsection{CNN Parameters}
Both networks were compiled with an Adam optimizer. After initial experimentation a learning rate of 0.01 was selected for the basic CNN, and 0.0005 for VGG-16, as this appeared to give the best performance in their respective architectures. Both used the mean absolute error (MAE) for their loss functions. MAE was selected as the loss function as we expected there to be significant level outliers in terms of their BC values, a feature better accounted for by calculating absolute errors rather than an average. 

Each CNN instance was given 100 epochs in which to train, though in the final experiments there were minimal improvements in terms of loss from as early as epoch 30 in certain configurations. Future implementations could usefully implement early stopping to avoid over-fitting and to save on computational resources.

\subsection{Level Selection and Splitting}

For each run 1000 levels were selected at random, distributed evenly between each generative approach for each game. For Mario they were evenly selected from the nine different generators, and for Boxoban they were evenly split between the three different sets.

For each combination of BC and CNN design a new instance of the network was generated and trained. For the CNN training, the 1000 levels were split into a set of 800 used to train the network, and a set of 200 to produce the final visualisation. 

We note that we are using a comparatively low amount of data to both train and test the neural networks. With just 1000 levels per run, a train/test split of 0.8 and a high of nine generators for the Mario domain this means that there are less than 90 levels from each generator in each training set. While this choice was primarily made to limit the computational resources required, it also helps to reinforce any potential claim about this approach having commercial utility. The more data that is required for a given generative space visualisation approach, the less widely useful it can be in domains where generating levels is either resource intensive, or in domains where it is a requirement to test numerous different configurations of a generator. 

\subsection{Number of Runs}

To increase the reliability of the results and to gain insight into how much variance there is with the approach, we conduct five runs for each of the two games and then present the average of the results for each compression method.

\subsection{Resources Required}

All experiments were run on a Dell laptop with an i5-10310U CPU with 16.0 GB of RAM. In total all experiments took $\approx$11 hours to run using this set up. 

\section{Results \& Discussion}\label{discussion}

\begin{table*}[h]
\centering
\caption{Main Results - Average of 5 Runs $\pm$ StdDev. '*' indicates cases where the average P value + StdDev across the 5 runs was greater than 0.01}
\label{table_mainresults}
\begin{tabular}{l|llll|llll|}
\cline{2-9}
                                 & \multicolumn{4}{c|}{Mario}                                                                                                                                                                                                                                                                                                                  & \multicolumn{4}{c|}{Boxoban}                                                                                                                                                                                                                                                                                                                \\ \cline{2-9} 
                                 & \multicolumn{1}{l|}{ES}                                                                & \multicolumn{1}{l|}{EC}                                                                & \multicolumn{1}{l|}{Lin}                                                              & Dens                                                              & \multicolumn{1}{l|}{ES}                                                               & \multicolumn{1}{l|}{CS}                                                               & \multicolumn{1}{l|}{ACS}                                                              & CC                                                                  \\ \hline
\multicolumn{1}{|l|}{VGG-16}     & \multicolumn{1}{l|}{\textbf{\begin{tabular}[c]{@{}l@{}}0.640\\ $\pm$0.0480\end{tabular}}} & \multicolumn{1}{l|}{\begin{tabular}[c]{@{}l@{}}0.255\\ $\pm$0.0437\end{tabular}}          & \multicolumn{1}{l|}{\begin{tabular}[c]{@{}l@{}}0.225\\ $\pm$0.0470\end{tabular}}         & \begin{tabular}[c]{@{}l@{}}0.166\\ $\pm$0.0455\end{tabular}          & \multicolumn{1}{l|}{\textbf{\begin{tabular}[c]{@{}l@{}}0.565\\ $\pm$0.303\end{tabular}}} & \multicolumn{1}{l|}{\textbf{\begin{tabular}[c]{@{}l@{}}0.555\\ $\pm$0.296\end{tabular}}} & \multicolumn{1}{l|}{\textbf{\begin{tabular}[c]{@{}l@{}}0.452\\ $\pm$0.252\end{tabular}}} & \begin{tabular}[c]{@{}l@{}}-0.00410\\ $\pm$0.0278*\end{tabular}        \\ \hline
\multicolumn{1}{|l|}{CNN-Basic}  & \multicolumn{1}{l|}{\begin{tabular}[c]{@{}l@{}}0.350\\ $\pm$0.154\end{tabular}}           & \multicolumn{1}{l|}{\begin{tabular}[c]{@{}l@{}}0.325\\ $\pm$0.115\end{tabular}}           & \multicolumn{1}{l|}{\textbf{\begin{tabular}[c]{@{}l@{}}0.527\\ $\pm$0.121\end{tabular}}} & \textbf{\begin{tabular}[c]{@{}l@{}}0.391\\ $\pm$0.0548\end{tabular}} & \multicolumn{1}{l|}{\begin{tabular}[c]{@{}l@{}}0.321\\ $\pm$0.147\end{tabular}}          & \multicolumn{1}{l|}{\begin{tabular}[c]{@{}l@{}}0.271\\ $\pm$0.136\end{tabular}}          & \multicolumn{1}{l|}{\begin{tabular}[c]{@{}l@{}}0.207\\ $\pm$0.115\end{tabular}}          & \textbf{\begin{tabular}[c]{@{}l@{}}0.0142\\ $\pm$0.0225*\end{tabular}} \\ \hline
\multicolumn{1}{|l|}{Vanilla DR} & \multicolumn{1}{l|}{\begin{tabular}[c]{@{}l@{}}0.506\\ $\pm$0.0340\end{tabular}}          & \multicolumn{1}{l|}{\textbf{\begin{tabular}[c]{@{}l@{}}0.337\\ $\pm$0.0165\end{tabular}}} & \multicolumn{1}{l|}{\begin{tabular}[c]{@{}l@{}}0.411\\ $\pm$0.0165\end{tabular}}         & \begin{tabular}[c]{@{}l@{}}0.246\\ $\pm$0.0130\end{tabular}          & \multicolumn{1}{l|}{\begin{tabular}[c]{@{}l@{}}0.0255\\ $\pm$0.0145\end{tabular}}        & \multicolumn{1}{l|}{\begin{tabular}[c]{@{}l@{}}0.0195\\ $\pm$0.0127\end{tabular}}        & \multicolumn{1}{l|}{\begin{tabular}[c]{@{}l@{}}-0.0248\\ $\pm$0.00930\end{tabular}}      & \begin{tabular}[c]{@{}l@{}}-0.0732\\ $\pm$0.00387\end{tabular}         \\ \hline
\end{tabular}
\end{table*}

\begin{figure}[h]
\centering
\includegraphics[width=3.2in]{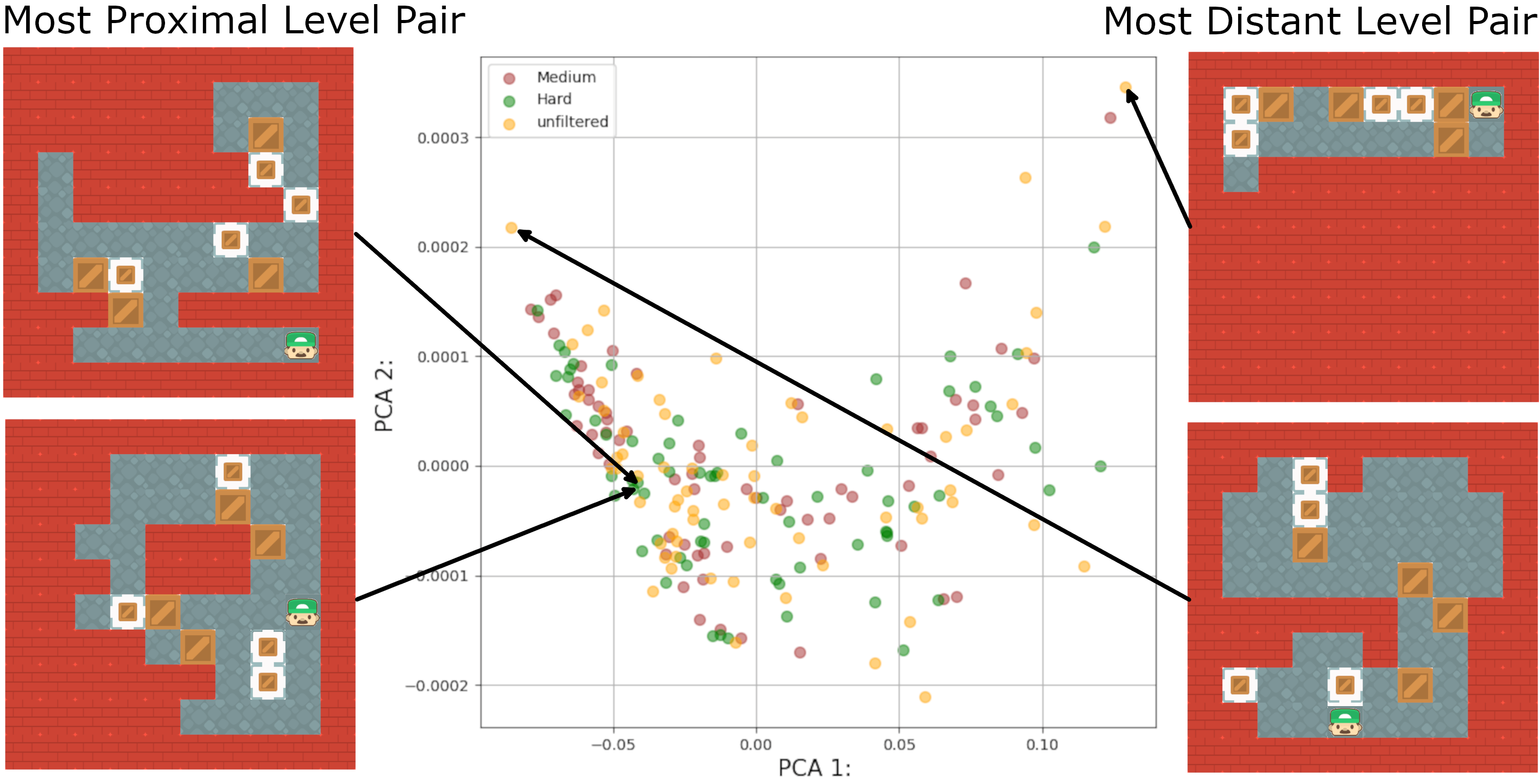}
\caption{Best performing Boxoban visualisation of all runs across all visualisation methods. From Run 2 of the VGG-16 CNN Implementation. Average BC correlation: 0.566}
\label{fig_bestboxoban}
\end{figure}

\begin{figure*}[h]
\centering
\includegraphics[width=6in]{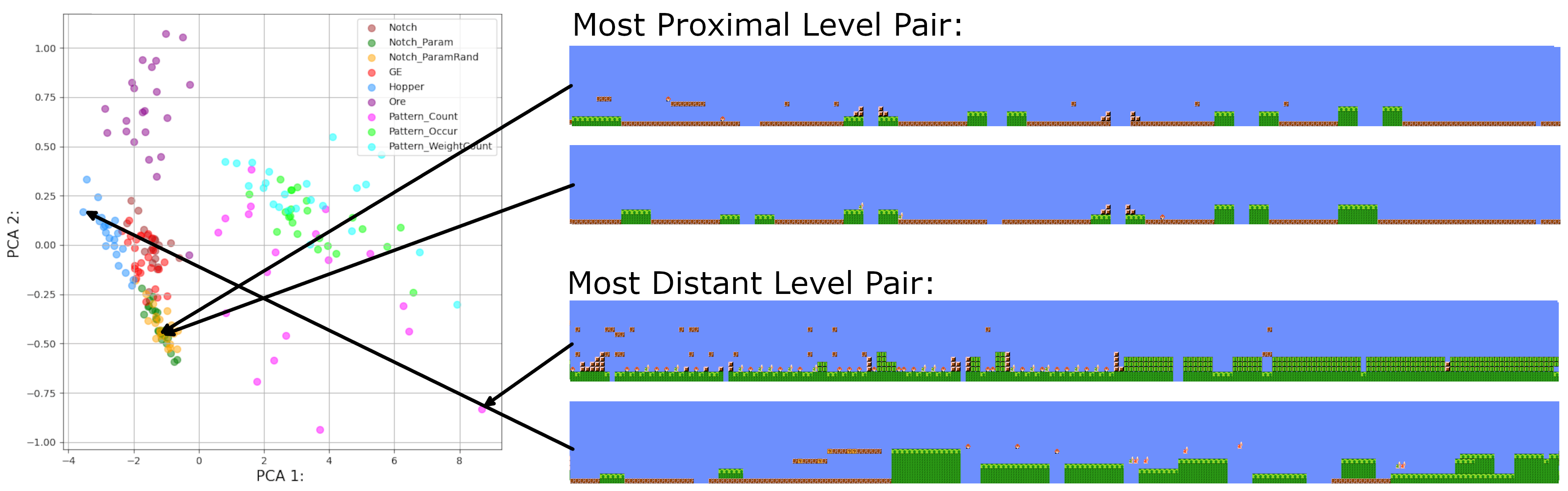}
\caption{Best performing Mario visualisation of all runs across all visualisation methods. From Run 2 of the Basic CNN Implementation. Average BC correlation: 0.476}
\label{fig_bestmario}
\end{figure*}

\subsection{Results Overview}

The experimental results give us a view of our approach that is both positive overall, but also mixed. The CNN-based visualisations substantially outperformed Vanilla DR in the Boxoban domain in terms of the BC correlations of their visualisations, and also performed well on individual BCs in the Mario domain. However, there was significant inconsistency in the performance of CNN approaches, suggesting that while the approach is promising, the implementation used needs significant refinement.

In Figures \ref{fig_bestboxoban} and \ref{fig_bestmario} show the best performing visualisations in terms of average BC correlation for the Boxoban and Mario domains respectively. These are presented to demonstrate what the final output of the visualisation process is, as well as to give an idea of how it could be used to qualitatively understand and compare generators. For example, the Mario visualisation suggests that the Ore generator highlighted in dark purple generates levels unlike any alternatives, whereas the Boxoban visualisation suggests that none of the three generative approaches produces substantially different levels to the other two. The fact that at least in the Mario domain the visualisation approach was able to separate and distinguish between the levels from alternative generators is positive, as characterising the output of alternative generators is a valuable potential use case.

Displayed with each visualisation is the pairs of levels which are the most proximal and distant in terms of their Euclidean distance in the visualisation. If the visualisation is working well, we would expect the most proximal pair to be the most similar pair of levels in the set and visa versa.  They suggest that while the Boxoban visualisations performed better than those for Mario in terms of the BC correlation heuristics, that this may not mean the visualisations are in fact more useful. If we examine the most proximal pair in Figure \ref{fig_bestboxoban}, it is not obvious that the levels are especially similar or that the experience of playing them would be similar either. In contrast the most proximal Mario pair in Figure \ref{fig_bestmario} do appear substantially similar. It is possible that the smaller a game level's representation is, the more important each tile choice becomes, and therefore the more important attaining a high level of BC correlation becomes. 

We expected due to its complexity and performance in other domains, that VGG-16 would outperform both the basic CNN and the Vanilla DR visualisations, and in 4/8 domains this was the case as we can see in Table \ref{table_mainresults}. Its performance was dominant in visualising the Boxoban domains, only struggling on the CC BC. However, given the underperformance across the board on the CC BC, we expect this to be an issue related to the structure of the CNN’s themselves, perhaps related to the convolutional window being too small to detect the corridor structure. The Basic CNN implementation also substantially outperformed Vanilla DR in the Boxoban domain, further reinforcing the idea that this CNN-based approach can perform well for certain games.

However, in the Mario domain the results were much more mixed. VGG-16 performed worse than the simple Basic CNN in 3 of 4 BCs, and Basic CNN was outperformed by Vanilla DR in 2 of 4. We suspect this failure is partly a result of quirks of the dataset. Vanilla DR found average EC correlations of 0.337, despite enemies being a sparse feature in most levels which should mean Vanilla DR is worse at accounting for them. This leads us to suspect that within the levels tested, EC correlates significantly with the easier to detect structural features like the amount of empty space. The poor performance of the CNN-based approaches is still concerning however, especially as they require significantly more set-up and compute costs than Vanilla DR. 

Additionally, the CNN-based approaches were prone to failure. Despite performing best in the Boxoban domain, the standard deviations for VGG-16 correlations were as high as 0.303 in the case of the ES BC. Examining the individual run data for both it and the Basic CNN shows that in some runs each CNN implementation failed during training to learn the relationship between level structure and BCs. Further experimentation is needed to investigate what was causing the intermittent failures, something we discuss in the next section.

\subsection{Future Work and Improvements}

On the technical front the primary aspect we need to address is the performance of the CNN training process. While we cannot be certain of what was causing the intermittent failures we have several different options to explore. First, we intend to further investigate the effect of tuning hyperparameters such as the learning rate to achieve a more consistent performance. We can also implement and evaluate alternative CNN implementations which have outperformed VGG-16 in certain domains, such as ResNet-50 \cite{he2016} and Xception \cite{chollet2017a}. However, all of these architectures share the conceptual weakness in that they were designed and optimised for analysing images rather than encoded game levels. Future work could investigate whether a bespoke CNN architecture for predicting level BCs could give improved results. 

Our future work will also further explore the performance gap between the Mario and Boxoban domains. It is possible that Vanilla DR is simply especially well suited to these specific Mario level sets, or that the CNN visualisation approach is ill suited to the larger level representations. Beyond further tuning the CNN training parameters, the best way of exploring this discrepancy is to explore alternative generative spaces, defined by alternative BCs. Related to this point, we also intend to start including BCs extracted from simulated play of the input levels by a level playing agent. If this approach produces correlations with these more complex BCs, this would be a cause for further optimism.

To make this approach more widely useful and accessible to game designers and researchers, we aim to investigate making the visualisation system publicly available in a form that could be quickly applied to new generators and new game domains. This could take the form of pre-trained CNNs embedded within a system that could be deployed to allow designers and researchers to produce visualisations without the need for training. It would also benefit from the development of an analytics layer, such as systems for displaying images of the outliers within a given generative space, or for giving representative samples of levels that are present within one generative space but not another. 
\section{Conclusion}\label{conclusion}
In this paper we introduced a novel approach for producing visualisations of the generative spaces of 2D level generators using trained CNNs, and conducted preliminary experiments using the approach on two level corpuses from Super Mario and Boxoban. The results indicated that it had the capacity to produce relatively robust results, but also that the current implementation is significantly prone to failure, as well as it under-performing compared to our benchmark visualisation approach in certain domains. As a result we find that while this approach has promise and warrants further investigation, more experimentation and tuning is required before it can be used reliably.

\section{Acknowledgements}
This work was supported by the EPSRC Centre for Doctoral Training in Intelligent Games \& Games Intelligence (IGGI) [EP/S022325/1]. We are also grateful for the helpful and informative feedback we received from the EXAG reviewers on the initial version of this work.

\bibliography{mainBibPureBibtex.bib}

\begin{thebibliography}{}

\bibitem[\protect\citeauthoryear{Alvarez \bgroup et al\mbox.\egroup
  }{2022}]{alvarez2022}
Alvarez, A.; Dahlskog, S.; Font, J.; and Togelius, J.
\newblock 2022.
\newblock Interactive {Constrained} {MAP}-{Elites}: {Analysis} and {Evaluation}
  of the {Expressiveness} of the {Feature} {Dimensions}.
\newblock {\em IEEE Transactions on Games} 14(2):202--211.

\bibitem[\protect\citeauthoryear{Awiszus, Schubert, and
  Rosenhahn}{2020}]{awiszus2020}
Awiszus, M.; Schubert, F.; and Rosenhahn, B.
\newblock 2020.
\newblock {TOAD}-{GAN}: {Coherent} style level generation from a single
  example.
\newblock In {\em Proceedings of the sixteenth {AAAI} conference on artificial
  intelligence and interactive digital entertainment}, {AIIDE}'20.
\newblock AAAI Press.
\newblock Number of pages: 7 tex.articleno: 2.

\bibitem[\protect\citeauthoryear{Ayesha, Hanif, and Talib}{2020}]{ayesha2020}
Ayesha, S.; Hanif, M.~K.; and Talib, R.
\newblock 2020.
\newblock Overview and comparative study of dimensionality reduction techniques
  for high dimensional data.
\newblock {\em Information Fusion} 59(Information Fusion):44--58.

\bibitem[\protect\citeauthoryear{Cerny~Green \bgroup et al\mbox.\egroup
  }{2020}]{cernygreen2020}
Cerny~Green, M.; Mugrai, L.; Khalifa, A.; and Togelius, J.
\newblock 2020.
\newblock Mario {Level} {Generation} {From} {Mechanics} {Using} {Scene}
  {Stitching}.
\newblock In {\em 2020 {IEEE} {Conference} on {Games} ({CoG})},  49--56.
\newblock Osaka, Japan: IEEE.

\bibitem[\protect\citeauthoryear{Chollet}{2017}]{chollet2017a}
Chollet, F.
\newblock 2017.
\newblock Xception: {Deep} {Learning} with {Depthwise} {Separable}
  {Convolutions}.
\newblock In {\em 2017 {IEEE} {Conference} on {Computer} {Vision} and {Pattern}
  {Recognition} ({CVPR})},  1800--1807.
\newblock Honolulu, HI: IEEE.

\bibitem[\protect\citeauthoryear{Cook \bgroup et al\mbox.\egroup
  }{2021}]{cook2021}
Cook, M.; Gow, J.; Smith, G.; and Colton, S.
\newblock 2021.
\newblock Danesh: {Interactive} {Tools} {For} {Understanding} {Procedural}
  {Content} {Generators}.
\newblock {\em IEEE Transactions on Games}.

\bibitem[\protect\citeauthoryear{Fontaine \bgroup et al\mbox.\egroup
  }{2021}]{fontaine2021}
Fontaine, M.~C.; Liu, R.; Khalifa, A.; Togelius, J.; Hoover, A.~K.; and
  Nikolaidis, S.
\newblock 2021.
\newblock Illuminating {Mario} {Scenes} in the {Latent} {Space} of a
  {Generative} {Adversarial} {Network}.
\newblock In {\em {AAAI}}.

\bibitem[\protect\citeauthoryear{Gardini \bgroup et al\mbox.\egroup
  }{2021}]{gardini2021}
Gardini, E.; Ferrarotti, M.~J.; Cavalli, A.; and Decherchi, S.
\newblock 2021.
\newblock Using {Principal} {Paths} to {Walk} {Through} {Music} and {Visual}
  {Art} {Style} {Spaces} {Induced} by {Convolutional} {Neural} {Networks}.
\newblock {\em Cognitive Computation} 13(2):570--582.

\bibitem[\protect\citeauthoryear{Guez \bgroup et al\mbox.\egroup
  }{2018}]{boxobanlevels}
Guez, A.; Mehdi, M.; Gregor, K.; Kabra, R.; Racaniere, S.; Weber, T.; Raposo,
  D.; Santoro, A.; Orseau, L.; Eccles, T.; Wayne, G.; Silver, D.; and
  Lillicrap, T.
\newblock 2018.
\newblock An investigation of {Model}-free planning: boxoban levels.

\bibitem[\protect\citeauthoryear{Guez \bgroup et al\mbox.\egroup
  }{2019}]{guez2019}
Guez, A.; Mirza, M.; Gregor, K.; Kabra, R.; Racanière, S.; Weber, T.; Raposo,
  D.; Santoro, A.; Orseau, L.; Eccles, T.; and {others}.
\newblock 2019.
\newblock An investigation of model-free planning.
\newblock In {\em International {Conference} on {Machine} {Learning}},
  2464--2473.
\newblock PMLR.

\bibitem[\protect\citeauthoryear{Guzdial \bgroup et al\mbox.\egroup
  }{2019}]{guzdial2019}
Guzdial, M.; Liao, N.; Chen, J.; Chen, S.-Y.; Shah, S.; Shah, V.; Reno, J.;
  Smith, G.; and Riedl, M.~O.
\newblock 2019.
\newblock Friend, {Collaborator}, {Student}, {Manager}: {How} {Design} of an
  {AI}-{Driven} {Game} {Level} {Editor} {Affects} {Creators}.
\newblock In {\em Proceedings of the 2019 {CHI} {Conference} on {Human}
  {Factors} in {Computing} {Systems}},  1--13.
\newblock Glasgow Scotland Uk: ACM.

\bibitem[\protect\citeauthoryear{He \bgroup et al\mbox.\egroup }{2016}]{he2016}
He, K.; Zhang, X.; Ren, S.; and Sun, J.
\newblock 2016.
\newblock Deep {Residual} {Learning} for {Image} {Recognition}.
\newblock In {\em 2016 {IEEE} {Conference} on {Computer} {Vision} and {Pattern}
  {Recognition} ({CVPR})},  770--778.
\newblock Las Vegas, NV, USA: IEEE.

\bibitem[\protect\citeauthoryear{Hervé and Salge}{2021}]{herve2021}
Hervé, J.-B., and Salge, C.
\newblock 2021.
\newblock Comparing {PCG} {Metrics} with {Human} {Evaluation} in {Minecraft}
  {Settlement} {Generation}.
\newblock In {\em The 16th {International} {Conference} on the {Foundations} of
  {Digital} {Games} ({FDG}) 2021}, {FDG}'21.
\newblock New York, NY, USA: Association for Computing Machinery.
\newblock event-place: Montreal, QC, Canada.

\bibitem[\protect\citeauthoryear{Horn \bgroup et al\mbox.\egroup
  }{2014}]{horn2014}
Horn, B.; Dahlskog, S.; Shaker, N.; Smith, G.; and Togelius, J.
\newblock 2014.
\newblock A comparative evaluation of procedural level generators in the
  {Mario} {AI} framework.
\newblock In Mateas, M.; Barnes, T.; and Bogost, I., eds., {\em Proceedings of
  the 9th {International} {Conference} on the {Foundations} of {Digital}
  {Games}, {FDG} 2014, {Liberty} of the {Seas}, {Caribbean}, {April} 3-7,
  2014}.
\newblock Society for the Advancement of the Science of Digital Games.

\bibitem[\protect\citeauthoryear{Irfan, Zafar, and Hassan}{2019}]{irfan2019}
Irfan, A.; Zafar, A.; and Hassan, S.
\newblock 2019.
\newblock Evolving {Levels} for {General} {Games} {Using} {Deep}
  {Convolutional} {Generative} {Adversarial} {Networks}.
\newblock In {\em 2019 11th {Computer} {Science} and {Electronic} {Engineering}
  ({CEEC})},  96--101.
\newblock Colchester, United Kingdom: IEEE.

\bibitem[\protect\citeauthoryear{Jadhav and Guzdial}{2021}]{jadhav2021a}
Jadhav, M., and Guzdial, M.
\newblock 2021.
\newblock Tile embedding: {A} general representation for level generation.
\newblock {\em Proceedings of the AAAI Conference on Artificial Intelligence
  and Interactive Digital Entertainment} 17(1):34--41.

\bibitem[\protect\citeauthoryear{Jemmali \bgroup et al\mbox.\egroup
  }{2020}]{jemmali2020}
Jemmali, C.; Ithier, C.; Cooper, S.; and El-Nasr, M.
\newblock 2020.
\newblock Grammar based modular level generator for a programming puzzle game.
\newblock {\em AAAI Conference on Artificial Intelligence and Interactive
  Digital Entertainment (AIIDE)}.

\bibitem[\protect\citeauthoryear{Khalifa and Togelius}{2009}]{khalifa2009}
Khalifa, A., and Togelius, J.
\newblock 2009.
\newblock Mario {AI} {Benchmark} -
  https://github.com/amidos2006/{Mario}-{AI}-{Framework}.

\bibitem[\protect\citeauthoryear{Racanière \bgroup et al\mbox.\egroup
  }{2017}]{racaniere2017}
Racanière, S.; Weber, T.; Reichert, D.~P.; Buesing, L.; Guez, A.; Rezende, D.;
  Badia, A.~P.; Vinyals, O.; Heess, N.; Li, Y.; Pascanu, R.; Battaglia, P.;
  Hassabis, D.; Silver, D.; and Wierstra, D.
\newblock 2017.
\newblock Imagination-{Augmented} {Agents} for {Deep} {Reinforcement}
  {Learning}.
\newblock In {\em Proceedings of the 31st {International} {Conference} on
  {Neural} {Information} {Processing} {Systems}}, {NIPS}'17,  5694--5705.
\newblock Red Hook, NY, USA: Curran Associates Inc.
\newblock event-place: Long Beach, California, USA.

\bibitem[\protect\citeauthoryear{Russakovsky \bgroup et al\mbox.\egroup
  }{2015}]{russakovsky2015a}
Russakovsky, O.; Deng, J.; Su, H.; Krause, J.; Satheesh, S.; Ma, S.; Huang, Z.;
  Karpathy, A.; Khosla, A.; Bernstein, M.; Berg, A.~C.; and Fei-Fei, L.
\newblock 2015.
\newblock {ImageNet} {Large} {Scale} {Visual} {Recognition} {Challenge}.
\newblock {\em International Journal of Computer Vision} 115(3):211--252.

\bibitem[\protect\citeauthoryear{Sarkar, Yang, and Cooper}{2020}]{sarkar2020}
Sarkar, A.; Yang, Z.; and Cooper, S.
\newblock 2020.
\newblock Controllable {Level} {Blending} between {Games} using {Variational}
  {Autoencoders}.
\newblock {\em arXiv:2002.11869 [cs]}.
\newblock arXiv: 2002.11869.

\bibitem[\protect\citeauthoryear{Simonyan and Zisserman}{2015a}]{simonyan2015a}
Simonyan, K., and Zisserman, A.
\newblock 2015a.
\newblock Very deep convolutional networks for large-scale image recognition.
\newblock {\em CoRR} abs/1409.1556.

\bibitem[\protect\citeauthoryear{Simonyan and Zisserman}{2015b}]{simonyan2015}
Simonyan, K., and Zisserman, A.
\newblock 2015b.
\newblock Very {Deep} {Convolutional} {Networks} for {Large}-{Scale} {Image}
  {Recognition}.
\newblock arXiv:1409.1556 [cs].

\bibitem[\protect\citeauthoryear{Smith and Whitehead}{2010}]{smith2010}
Smith, G., and Whitehead, J.
\newblock 2010.
\newblock Analyzing the expressive range of a level generator.
\newblock In {\em Proceedings of the 2010 {Workshop} on {Procedural} {Content}
  {Generation} in {Games} - {PCGames} '10},  1--7.
\newblock Monterey, California: ACM Press.

\bibitem[\protect\citeauthoryear{Smith, Padget, and Vidler}{2018}]{smith2018}
Smith, T.; Padget, J.; and Vidler, A.
\newblock 2018.
\newblock Graph-based generation of action-adventure dungeon levels using
  answer set programming.
\newblock In {\em Proceedings of the 13th {International} {Conference} on the
  {Foundations} of {Digital} {Games}},  1--10.
\newblock Malmö Sweden: ACM.

\bibitem[\protect\citeauthoryear{Summerville}{2018}]{summerville2018}
Summerville, A.
\newblock 2018.
\newblock Expanding {Expressive} {Range}: {Evaluation} {Methodologies} for
  {Procedural} {Content} {Generation}.
\newblock In {\em Fourteenth {Artificial} {Intelligence} and {Interactive}
  {Digital} {Entertainment} {Conference}}.

\bibitem[\protect\citeauthoryear{Togelius, Karakovskiy, and
  Baumgarten}{2010}]{togelius2009MarioAI2010}
Togelius, J.; Karakovskiy, S.; and Baumgarten, R.
\newblock 2010.
\newblock The 2009 {Mario} {AI} {Competition}.
\newblock In {\em {IEEE} {Congress} on {Evolutionary} {Computation}},  1--8.
\newblock Barcelona, Spain: IEEE.

\bibitem[\protect\citeauthoryear{Volz \bgroup et al\mbox.\egroup
  }{2018}]{volz2018a}
Volz, V.; Schrum, J.; Liu, J.; Lucas, S.~M.; Smith, A.; and Risi, S.
\newblock 2018.
\newblock Evolving mario levels in the latent space of a deep convolutional
  generative adversarial network.
\newblock In {\em Proceedings of the {Genetic} and {Evolutionary} {Computation}
  {Conference}},  221--228.
\newblock Kyoto Japan: ACM.

\bibitem[\protect\citeauthoryear{Withington and
  Tokarchuk}{2022}]{withington2022}
Withington, O., and Tokarchuk, L.
\newblock 2022.
\newblock Compressing and {Comparing} the {Generative} {Spaces} of {Procedural}
  {Content} {Generators}.
\newblock arXiv:2205.15133 [cs].

\bibitem[\protect\citeauthoryear{Withington}{2020}]{withington2020}
Withington, O.
\newblock 2020.
\newblock Illuminating super mario bros: quality-diversity within platformer
  level generation.
\newblock In {\em Proceedings of the 2020 {Genetic} and {Evolutionary}
  {Computation} {Conference} {Companion}},  223--224.
\newblock Cancún Mexico: ACM.

\bibitem[\protect\citeauthoryear{Wulff-Jensen \bgroup et al\mbox.\egroup
  }{2018}]{wulff-jensen2018}
Wulff-Jensen, A.; Rant, N.~N.; Møller, T.~N.; and Billeskov, J.~A.
\newblock 2018.
\newblock Deep {Convolutional} {Generative} {Adversarial} {Network} for
  {Procedural} {3D} {Landscape} {Generation} {Based} on {DEM}.
\newblock In Brooks, A.~L.; Brooks, E.; and Vidakis, N., eds., {\em
  Interactivity, {Game} {Creation}, {Design}, {Learning}, and {Innovation}},
  volume 229. Cham: Springer International Publishing.
\newblock  85--94.
\newblock Series Title: Lecture Notes of the Institute for Computer Sciences,
  Social Informatics and Telecommunications Engineering.

\end{thebibliography}
\bibliographystyle{aaai}

\end{document}